\documentclass[twocolumn,aps,10pt,prl]{revtex4-1}

\def\mysections#1{{\bf #1.} }

\usepackage[dvips]{graphicx}
\usepackage{epsfig,amsmath,amssymb,verbatim,mathrsfs,array,layout,textcomp,latexsym, slashed,graphicx,bbm}

\usepackage[normalem]{ulem}
\usepackage{appendix}

\usepackage{rotating}
\usepackage{hyperref}
\usepackage[usenames,dvipsnames]{color}

\usepackage{soul}

\renewcommand\){\right)}

\newcommand{\be}{\begin{equation}}
\newcommand{\ee}{\end{equation}}
\newcommand{\bea}{\begin{eqnarray}}
\newcommand{\eea}{\end{eqnarray}}

\definecolor{gbcolor}{rgb}{.8,.3,.1}
\definecolor{gbcolor2}{rgb}{.8,.1,.7}

\def\beq{\begin{equation}}
\def\eeq{\end{equation}}

\newcommand{\eq}[1]{(\ref{#1})}

\begin{document}

\begin{flushleft}                          
\footnotesize DESY 16-049, IPPP/16/25
\end{flushleft}

\title{Unifying inflation with the axion, dark matter, \\baryogenesis and the seesaw mechanism}

\author{Guillermo Ballesteros$^{1}$}\email{guillermo.ballesteros@cea.fr}
\author{Javier Redondo$^{2,3}$}\email{jredondo@unizar.es}
\author{Andreas Ringwald$^{4}$}\email{andreas.ringwald@desy.de}
\author{Carlos Tamarit$^5$}\email{carlos.tamarit@durham.ac.uk}
\affiliation{$^1$Institut de Physique Th\'eorique, Universit\'e Paris Saclay, CEA, CNRS. 91191 Gif-sur-Yvette, France}
\affiliation{$^2$Departamento de F\'isica Te\'orica, Universidad de Zaragoza. Pedro Cerbuna 12, E-50009, Zaragoza, Spain}
\affiliation{$^3$Max-Planck-Institut f\"ur Physik, F\"ohringer Ring 6, 80805 M\"unchen, Germany}
\affiliation{$^4$DESY, Notkestr. 85, 22607 Hamburg, Germany}
\affiliation{$^5$Institute for Particle Physics Phenomenology, Durham University South Road, DH1 3LE, United Kingdom}

\begin{abstract}

A minimal extension of the Standard Model (SM) {with a single new mass scale and} providing a complete and consistent picture of particle physics and cosmology
up to  the Planck scale is presented. We add to the SM
three right-handed  SM-singlet neutrinos, a new vector-like color triplet fermion and a complex SM singlet scalar $\sigma$ {that stabilises the Higgs potential and} whose vacuum expectation value at $\sim 10^{11}$\,GeV breaks lepton number and  a Peccei-Quinn symmetry simultaneously. Primordial inflation is produced by a combination of $\sigma$ {(non-minimally coupled to the scalar curvature)} and the SM Higgs. Baryogenesis proceeds via thermal leptogenesis. At low energies, the model reduces to the SM, augmented by seesaw-generated neutrino masses, plus the axion, which solves the strong CP problem and accounts for the dark matter in the {Universe}. The model predicts a minimum value of the tensor-to-scalar ratio {$r\simeq 0.004$,} {running of the {scalar} spectral index $\alpha\simeq - {7\times} 10^{-4}$, the axion mass $m_A\sim 100\,\mu{\rm eV}$ and} {cosmic axion background radiation corresponding to 
an increase of the effective number of relativistic neutrinos} {of $\sim 0.03$. It} can be probed decisively by the next generation of cosmic microwave
background and axion dark matter experiments.

\end{abstract}

\maketitle

\section{Introduction}
The Standard Model of particle physics (SM) describes with exquisite precision the interactions of all known elementary particles. In spite of intensive searches, no significant deviation from the SM has been detected in collider or other particle physics experiments {\cite{Olive:2016xmw}.} However, several long-standing problems  indicate that new physics beyond the SM is needed to achieve a complete description of Nature.  First of all, there is overwhelming evidence, ranging from the cosmic microwave background (CMB) to the shapes of the rotation curves of spiral galaxies, that nearly 26\% of the Universe is made of yet unidentified dark matter (DM) \cite{Ade:2015xua}. Moreover, the SM cannot generate the primordial inflation needed to solve the horizon and flatness problems of the Universe, as well as to explain the statistically isotropic, Gaussian and nearly scale invariant fluctuations of the CMB \footnote{If the Higgs field has a large coupling to the curvature $R$, inflation might be obtained \cite{Bezrukov:2007ep}. However, as we will argue, this requires physics beyond the SM.}. The SM also lacks enough CP violation to explain why the Universe contains a larger fraction of baryonic matter than of anti-matter. Aside from these three problems at the interface between particle physics and cosmology, the SM suffers from a variety of intrinsic naturalness issues. In particular, the neutrino masses are disparagingly smaller than any physics scale in the SM and, similarly, the strong CP problem states that the $\theta$-parameter of quantum chromodynamics (QCD) is constrained from measurements of the neutron electric dipole moment {\cite{Afach:2015sja,Schmidt-Wellenburg:2016nfv}} to lie below  an unexpectedly small value: {$|\theta|\lesssim 10^{-10}$.} 

In this Letter we show that these problems may be intertwined in a remarkably simple way, with a solution pointing to a unique new physics scale around $10^{11}$ GeV. The SM extension we consider consists just of {a KSVZ-like} axion model \cite{Kim:1979if,Shifman:1979if} and three right-handed (RH) heavy SM-singlet neutrinos \footnote{One may also chose alternatively the DFSZ axion model. The inflationary predictions in this model stay the same, but the window in the axion mass will move to larger values \cite{Kawasaki:2014sqa}. Importantly, in this case the PQ symmetry is required to be an accidental rather than an exact symmetry in order to avoid the overclosure of the Universe due to domain walls  \cite{Ringwald:2015dsf}}. This extra matter content was recently proposed in \cite{Dias:2014osa}, where it was emphasised that in addition to  solving the strong CP problem, providing a good dark matter candidate (the axion), explaining the origin of the small SM neutrino masses (through an induced seesaw mechanism) and the baryon asymmetry of the Universe (via thermal leptogenesis), it could also stabilise the effective potential of the SM at high energies thanks to a threshold mechanism \cite{Lebedev:2012zw,EliasMiro:2012ay}. This extension also leads to successful primordial inflation by using the modulus of the KSVZ SM singlet scalar field \cite{Fairbairn:2014zta}.  Adding a cosmological constant to account for the present acceleration of the Universe, this Standard Model Axion Seesaw Higgs portal inflation (SMASH) model offers  a self-contained description of particle physics from the electroweak scale to the Planck scale and of cosmology from inflation until today. Although some parts of our SMASH model have been considered separately \cite{Langacker:1986rj,Shin:1987xc,Shaposhnikov:2006xi,Bezrukov:2007ep,Lerner:2009xg,Lebedev:2011aq,Fairbairn:2014zta,Boucenna:2014uma,Bertolini:2014aia,Salvio:2015cja,Kahlhoefer:2015jma,Clarke:2015bea,Ahn:2015pia}, a model incorporating all of them simultaneously had not been proposed {until now.} 
{Remarkably,  SMASH} {can accommodate} {the} {constraints from  cosmological observations and Higgs stability, successfully reheat the Universe, provide the correct dark matter abundance and explain the origin of the baryon asymmetry.}
{In this Letter, we present the most important aspects and predictions of SMASH}. {Further details are given in \cite{Ballesteros:2016xej}.}

\section{The SMASH model}
We extend the SM with a new complex singlet scalar field $\sigma$ {and a Dirac fermion $\mathcal{Q}$, which can be split in} two Weyl fermions $Q$ and $\tilde Q $ in the $\mathbf 3$ and $\mathbf{\bar{3}}$ representations of $SU(3)_c$ {with charges} $-1/3$ and $1/3$ under $U(1)_Y$. {This ensures that $\mathcal{Q}$ can coannihilate and decay into SM} {quarks, thereby evading possible overabundance problems~\cite{Nardi:1990ku,Berezhiani:1992rk}.}
We also add three RH fermions $N_i$. 
{The model is endowed with a new Peccei-Quinn (PQ) global $U(1)$ symmetry \cite{Peccei:1977hh}, which also plays the role of lepton number {in our case.}  {Using left-handed Weyl spinors, we denote by $q_i$, $u_i$ and $d_i$ the SM quark doublet and the conjugates of the right-handed quarks of each generation $i=1,2,3$;  and by $L_i$ and $E_i$ the corresponding  lepton doublet and the conjugate of the right-handed lepton. Denoting the Higgs by $H$,} {the charges under the PQ symmetry are: $q(1/2)$, $u(-1/2)$, $d(-1/2)$, $L(1/2)$, $N(-1/2)$, $E(-1/2)$, $Q(-1/2)$, $\tilde Q(-1/2)$, $\sigma(1)$, $H(0)$.} {The most general Yukawa couplings involving the new fields are:} {${\cal L}\supset -[F_{ij}L_i\epsilon H N_j+\frac{1}{2}Y_{ij}\sigma N_i  N_j+y\, \tilde Q \sigma Q+z_i\,\sigma Q d_i+h.c.]$, where $\epsilon$ is the two-component antisymmetric symbol.}  {The Yukawa couplings $F$ and $Y$} realise the seesaw mechanism once $\sigma$ acquires a vacuum expectation value (VEV) {$\langle\sigma\rangle=v_\sigma/\sqrt{2}$,} giving a neutrino mass matrix of the form {$m_\nu=-FY^{-1}F^Tv^2/(\sqrt{2}v_\sigma)$,} with $v=246$ GeV. The strong CP problem is solved as in the standard KSVZ scenario, with the role of the axion decay constant,  $f_A$, played by $v_\sigma=f_A$. Due to non-perturbative QCD effects, the angular part of {$\sigma=(\rho+v_\sigma)\exp(iA/f_A)/\sqrt{2}$,} the axion field $A$ \cite{Weinberg:1977ma,Wilczek:1977pj}, gains a potential with an absolute minimum at $A=0$. At energies above the QCD scale, the axion-gluon coupling is $\mathcal{L}\supset -(\alpha_s/8\pi)(A/f_A) G\tilde G$, solving the strong CP problem when 
$\langle A\rangle$ relaxes to zero\footnote{{Since the $\tilde Q,Q$ quarks have hypercharge and the PQ charge assignments are different than in the standard KSVZ scenario, the axion has a non-standard coupling to the photon, as well as a coupling to neutrinos \cite{Ballesteros:2016xej}}}. The latest lattice computation of the axion mass gives $m_A=(57.2\pm0.7)(10^{11}{\rm GeV}/f_A)\,\mu{\rm eV} $~\cite{Borsanyi:2016ksw}.

\section{Inflation}
{Given the symmetries of SMASH, the most general renormalisable tree-level potential is}
\begin{align}
\label{scalar_potential} \nonumber
V(H,\sigma ) & = \lambda_H \left( H^\dagger H - \frac{v^2}{2}\right)^2
+\lambda_\sigma \left( |\sigma |^2 - \frac{v_{\sigma}^2}{2}\right)^2\\ &+
2\lambda_{H\sigma} \left( H^\dagger H - \frac{v^2}{2}\right) \left( |\sigma |^2 - \frac{v_{\sigma}^2}{2}\right)\,.
\end{align}

\begin{figure}[t]
\begin{center}
\includegraphics[width=0.475\textwidth]{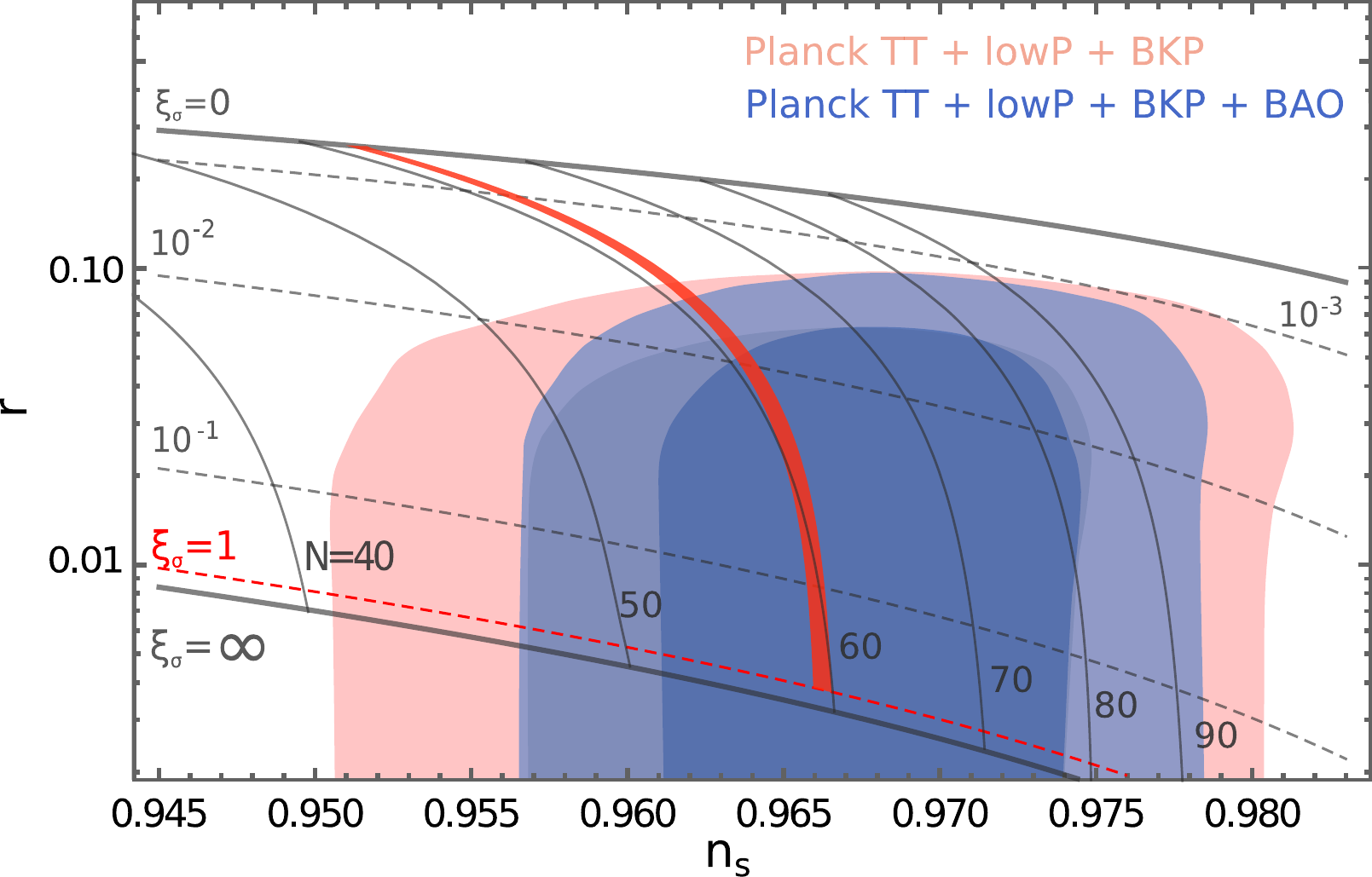}
\caption{\label{fig:r_vs_ns}  \small {The tensor-to-scalar ratio, $r$, vs the scalar spectral index, $n_s$, at} {$k_0=0.002$}  {Mpc$^{-1}$ for the inflationary potential \eq{genpotential}, assuming} {$|\lambda_{H\sigma}|\ll\lambda_H$.} {We show lines of constant $\xi_\sigma$ (dashed) and constant number of e-folds from the time the scale} {$k_0=0.002$} {Mpc$^{-1}$ exits the horizon to the end of inflation  (thin solid). 
In SMASH, the EOS of the Universe is like $w=1/3$ immediately after inflation, which allows to predict $N$ (thick red line).}
Coloured regions show observational constraints at 68\% and 95\% CL from \cite{Ade:2015lrj}.}
\end{center}
\end{figure}

In the unitary gauge, there are two scalar  fields that could  drive inflation: $h$, the neutral component of the Higgs doublet $H^t=(0\,,h)/\sqrt{2}$, and the modulus of the new singlet, {$\rho^2=2\,|\sigma|^2$.} In the context of the SM,  it was proposed in \cite{Bezrukov:2007ep}  that $h$ could be the inflaton if it is non-minimally coupled to the scalar curvature $R$ through a term $\mathcal{L}\supset -\sqrt{-g}\,\xi_H\,H^\dagger H\,R$ \cite{Salopek:1988qh}, with $\xi_H\sim 10^4$. Such a  large value of $\xi_H$ is required by the constraint $\xi_H\sim 10^5 \sqrt{\lambda_H}$ to fit the amplitude of primordial fluctuations and it implies that perturbative unitarity breaks down at the scale {$\Lambda_U=M_P/\xi_H\ll M_P$} \cite{Burgess:2009ea, Barbon:2009ya}, where $M_P=1/\sqrt{8\pi\,G}$ is the reduced Planck mass. This raises a serious difficulty for Higgs inflation, which requires Planckian values of $h$ and an energy density of order $\Lambda_U^2$. Since  new physics is expected at or below $\Lambda_U$ to restore unitarity, the predictivity of Higgs inflation is lost, because the effect of this new physics on inflation is undetermined. This issue affects some completions of the SM such as the $\nu$MSM \cite{Asaka:2005an,Asaka:2005pn}  and the model proposed in \cite{Salvio:2015cja}. 
Instead, inflation in SMASH is mostly driven by $\rho$, with a non-minimal coupling $\mathcal{L}\supset -\sqrt{-g}\,\xi_\sigma\,\sigma^* \sigma\,R$, {where $\xi_\sigma\lesssim 1$ ensures that the scale of perturbative unitarity breaking is at $M_P$ (provided that also $\xi_H\lesssim 1$).}
Neglecting $\xi_H$ \footnote{{Taking into account radiative corrections to $\xi_H$ and $\xi_\sigma$ one can check that the window $2\times 10^{-3}\lesssim\xi_\sigma\lesssim 1$ ensures that $\xi_H$ can be neglected with respect to $\xi_\sigma$}}, predictive slow-roll inflation in SMASH can happen along two directions in field space: the  $\rho$-direction for $\lambda_{H\sigma}>0$ and the {line} $h/\rho=\sqrt{-\lambda_{H\sigma}/\lambda_H}$ for $\lambda_{H\sigma}<0$. We call them hidden scalar inflation (HSI) and Higgs-hidden scalar inflation (HHSI), {respectively. In both cases, inflation} can be described {in the Einstein frame}  by a single canonically normalised field $\chi$ with potential
\be
\label{genpotential}
\tilde V(\chi) = \frac{\lambda }{4}\rho(\chi)^4\left(1+\xi_{\sigma}\frac{\rho(\chi)^2}{M_P^2}\right)^{-2}\,,
\ee 
where $\lambda$ {stands for $\lambda_\sigma$ in HSI and for  
 $\tilde\lambda_\sigma=\lambda_\sigma-\lambda_{H\sigma}^2/\lambda_{H}$ in HHSI. }
The field $\chi$ is the solution of $\Omega^2\,d\chi/d\rho\simeq (b\,\Omega^2+6\,\xi_\sigma^2\,\rho^2/M_P^2)^{1/2}$, {being} $\Omega\simeq 1+\xi_\sigma\,\rho^2/M_P^2$ the Weyl transformation into the Einstein frame; and {$b=1$ (for HSI) or $b=1+|\lambda_{H\sigma}/\lambda_H|$ (for HHSI). The small value of $|\lambda_{H\sigma}|$ required for stability (see below) typically means that $b\sim1$ in HHSI, which makes impossible distinguishing in practice between HSI and HHSI from the inflationary potential. However, even a small Higgs component in the inflaton is relevant for reheating, as we will later discuss. The predictions of the potential \eq{genpotential}} in the case $\lambda=\lambda_\sigma$ {(or $b\rightarrow 1$ in HHSI)} for { the tensor-to-scalar ratio $r$} vs the scalar spectral index $n_s$ are shown in FIG. \ref{fig:r_vs_ns} for various values of $\xi_\sigma$. 

{In SMASH, the equation of state} {(EOS)} {of the Universe after inflation is $w=1/3$  (like radiation) uninterruptedly until the} {standard epoch of matter-radiation equality is reached; see the reheating section below.} {This allows to compute the number of e-folds of inflation, $N(k)$, for any comoving scale, $k$, matching precisely the predictions for the inflationary spectrum with the observations of the CMB \cite{Liddle:2003as}. {This} determines the {thick line} of FIG.\ \ref{fig:r_vs_ns} as the SMASH prediction for $r(n_s)$ and {$N(k_0)$} at the {fiducial scale $k_0 \equiv 0.002$ Mpc$^{-1}$, which we use through the Letter for all the primordial inflationary parameters.}
The prediction spans {$N\sim(59,62)$}, depending on $n_s$, and its width {($\sim 1$ e-fold)} {quantifies the small} uncertainty on the transient regime from the end of inflation to {radiation domination.}}

{Note that the the} {condition $\xi_\sigma\lesssim 1$ corresponds to {$r\gtrsim  0.004$,} {which is within the planned sensitivities of {PIXIE \cite{Kogut:2011xw}, LiteBird \cite{Matsumura:2013aja}, CMB-S4 \cite{Abazajian:2016yjj} and COrE+ (which will measure $r$ with an error of $\Delta r\sim 4\times 10^{-4}$).} 
The {joint} {constraints} of the Planck satellite and the BICEP/Keck array \cite{Ade:2015xua,Array:2015xqh} 
{give}  {$r< 0.07$ at 95\% CL,} {corresponding in SMASH to  $\xi_\sigma\gtrsim 6\times 10^{-3}$.} 
{Taking into account  the former constraints, }the spectral index $n_s$ {at $k=k_0$ lies in the interval $(0.962, 0.966)$, and its running $\alpha=d\, n_s/d \ln k$ {lies} in {the range} $(- 7,-6)\times 10^{-4}$, which may be} probed e.g.\ by future observations of the 21 cm emission line of Hydrogen \cite{Mao:2008ug}. {Since inflation is effectively single-field slow-roll, non-Gaussian features are suppressed by $\sim(1-n_s)$ \cite{Acquaviva:2002ud,Maldacena:2002vr}.} These values of the primordial parameters are perfectly compatible with the latest CMB data, and the amount of inflation that is produced solves the horizon and flatness problems. 
Given the current bounds on $r$ and $n_s$, {and the fact that fitting the amplitude of primordial scalar fluctuations requires $\xi_\sigma \sim 10^5 \sqrt{\lambda}$,} fully consistent (and predictive)  inflation in SMASH occurs if {$5\times 10^{-13}\lesssim \lambda \lesssim 5\times 10^{-10}$.}

\section{Stability}
For the measured central values of the Higgs and top quark masses {\cite{Olive:2016xmw},} the Higgs quartic coupling of the SM becomes negative at $h=\Lambda_I\sim 10^{11}$ GeV \footnote{$\Lambda_I$ is very sensitive to small variations of the top mass, to the extent that the potential may be completely stable for sufficiently low (but still allowed) values.}. If no new physics changes this behaviour, Higgs inflation is not viable, since it requires a positive potential at Planckian field values. Moreover, {this instability is} a problem even if another field drives inflation. This is because {scalars that are light compared} to the Hubble scale, $\mathcal H$, acquire fluctuations of order {$\sim\mathcal{H}/2\pi$.} These {can make the Higgs field move} into the instability region of the potential, {which would contradict} the present electroweak vacuum \footnote{The fluctuations can be suppressed if $\xi_H\gtrsim 1$, which induces a large effective mass for the Higgs during inflation.  In that case it would still be necessary to check that the classical trajectory of the fields does not fall in the negative region of the potential}. 
Remarkably, {the} Higgs portal term $\propto \lambda_{H\sigma}$ in \eq{scalar_potential} allows {stability of the SMASH potential} via the threshold-stabilisation mechanism {of \cite{Lebedev:2012zw,EliasMiro:2012ay}, {which relies on a} {nontrivial matching with the SM potential} {at low energies.} The {matched Higgs quartic in the SM is}  {$\tilde\lambda_H\equiv\lambda_H-\delta$, where the threshold correction is $\delta\equiv\lambda^2_{H\sigma}/\lambda_\sigma$. Even if the running of $\tilde\lambda_H$ in the SM makes it negative, the actual Higgs quartic coupling} {in the UV theory,} $\lambda_H$, {can remain positive provided that $\delta$ is large enough.}  } 
{A more detailed analysis \footnote{{At large field values, stability demands positivity of quartic couplings, but at intermediate field values, where negative quadratic interactions are important, one has to ensure 
positivity of the potential along the} {potential} {energy valleys}} shows that, for} $\lambda_{H\sigma}>0$, absolute stability requires \cite{Ballesteros:2016xej}
\bea
\label{stabilitycondition1}
\left\{\begin{array}{cc}  
\tilde \lambda_H, \tilde \lambda_\sigma>0, \quad\text{for}\quad h < \sqrt{2}\Lambda_h\\
         \lambda_H,         \lambda_\sigma>0, \quad\text{for}\quad h > \sqrt{2}\Lambda_h \\
\end{array}\right.,
\eea
{where $\Lambda_h^2\equiv\lambda_{H\sigma}\,v_\sigma^2/\lambda_H$ and all the couplings run with the {beta functions} of SMASH, not the SM. The scale $\sqrt{2}\Lambda_h$ arises as the divide between large and small field values of $h$, for which $v_\sigma$ cannot be neglected and the quadratic interactions are relevant, as can be seen from \eq{scalar_potential}.} Instead, for $\lambda_{H\sigma}<0$, the stability condition is just
$\tilde \lambda_H, \tilde \lambda_\sigma>0$, for all $h$. 
The Higgs direction is the one most prone to be destabilised (from top loops) and the potential must remain positive beyond the $h\sim M_P$ values needed for inflation. A one-loop analysis shows that a value of $\delta$ above $10^{-3}$--$10^{-1}$ (depending on the top mass, see FIG.\ \ref{fig:delta}) ensures stability up to $h_{30}\approxeq 30 M_P$ for a Higgs mass  of $125.09$~GeV. }{
Finally, in SMASH, instabilities could also originate in the {direction of $\rho$} due to quantum corrections from {$N_i$ and $Q,\tilde Q$. Stability in this direction, requires $6 y^4+\sum Y^4_{ii}\lesssim 16\pi^2\lambda_\sigma/\log\(h_{30}/\sqrt{2\lambda_\sigma}v_\sigma\)$} \footnote{{In this expression and in FIG.\ \ref{fig:delta}, we demand stability up to a large RG scale $\mu\sim h_{30}$, which is sufficiently higher than $h$ during inflation.}.}.}

\begin{figure}[t]
\begin{center}
\includegraphics[width=0.43\textwidth]{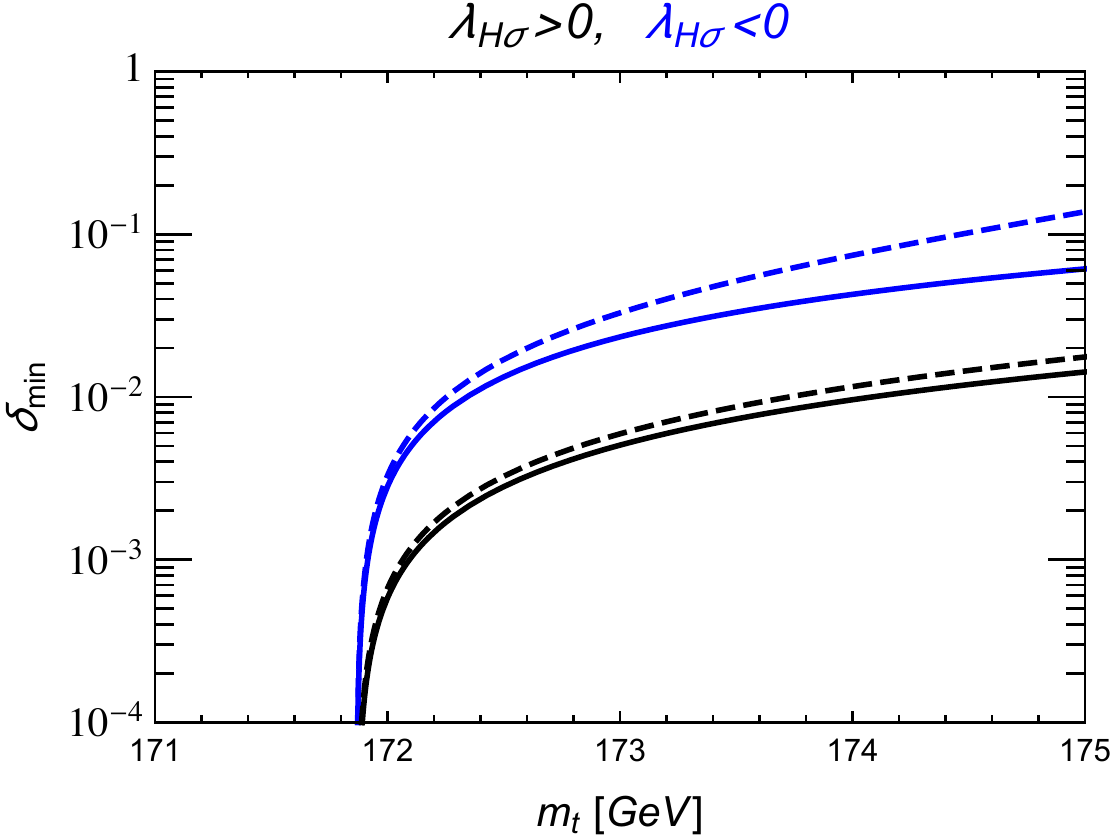}
\caption{\label{fig:delta}  \small Minimum value of the threshold correction to the Higgs quartic coupling, $\delta=\lambda_{H\sigma}^2/\lambda_{\sigma}$, for stable SMASH potentials at RG scales $\mu=m_\rho$ (solid) and $\mu=30 M_P$ (dashed), for $\lambda_{H\sigma}>0$ (black) and $\lambda_{H\sigma}<0$ (blue).}
\end{center}
\end{figure}

\section{Reheating} 
{SMASH provides a complete model of cosmology for which the evolution after inflation can be calculated. The PQ symmetry is spontaneously broken during inflation by the large evolving value of $\rho$.} Slow-roll inflation ends at  $\rho_{\rm end}\sim \mathcal{O}(M_P)$, where the effect of $\xi_\sigma$ is negligible. Since $\rho_{\rm end} \gg v_\sigma$, the inflaton starts to undergo Hubble-damped oscillations in a quartic potential. 

{The first oscillations of the inflaton constitute a phase of so-called preheating \cite{Kofman:1997yn}, during which} fluctuations of $\sigma$ in the direction orthogonal to the inflaton increase exponentially.} The post-inflationary background can be understood as {a} homogeneous condensate of particles with energy given by the oscillation frequency  $\omega(t)\sim\sqrt{\lambda}\rho_{\rm end}/a(t)$, {where $a(t)$ is the scale factor of the Universe and $t$ denotes cosmic time \footnote{We use natural units for which $\hbar=c=1$.}.} In SMASH, $\lambda$ is the weakest coupling and thus SM particles coupled to the inflaton have effective masses $\propto \rho(t)$, which are much larger than $\omega(t)$ except when $\rho(t)\sim 0$.  {Higgs particles and} electroweak bosons could in principle be produced by parametric resonance \cite{Greene:1997fu} {at these crossings}  but they either have large self-interactions or decay very efficiently into SM fermions.  In contrast, the effective mass of $\sigma$ excitations is $\sim \sqrt{\lambda}\rho(t)\sim \omega(t)$, {which allows them to grow} by parametric resonance.  The growth of fluctuations of a complex {inflaton field} in a quartic potential was studied {analytically in~\cite{Greene:1997fu} and} numerically in \cite{Tkachev:1998dc}. Our own numerical simulations~\cite{Ballesteros:2016xej} corroborate their results. {After the first $\sim 14$ oscillations after inflation,} {the fluctuations of $\sigma$ become} {as large as the inflaton amplitude $\langle|\sigma|^2\rangle  \sim \rho_{\rm end}^2/a^2$, so the PQ  symmetry is non-thermally restored.} Only if $v_\sigma$ were larger than $\sim 10^{-2}M_P$ would the field $\rho$ get trapped around its minimum $\rho=v_\sigma$ before the non-thermal restoration can occur. However, such high values of $v_\sigma$ are ruled out by CMB axion isocurvature constraints~\cite{Fairbairn:2014zta} \footnote{
Reference \cite{Fairbairn:2014zta} does not mention parametric resonance and non-thermal restoration of the PQ symmetry, because it {is} model dependent. The authors of \cite{Fairbairn:2014zta} showed that a large inflaton VEV suppresses the isocurvature constraints. We have redone the analysis finding stronger bounds~\cite{Ballesteros:2016xej}. 
Probably, ref. \cite{Fairbairn:2014zta} used the Jordan frame VEV instead of the effective axion
decay constant in the Einstein frame, $~\chi$. 
In any case, $v_\sigma \gtrsim 10^{-2}M_P$ is safely excluded. }. 

{Aside from these common features, reheating progresses differently for HSI and HHSI. The reason is that the small Higgs component of the inflaton in HHSI (which is lacking in HSI) accelerates in that case the production of SM particles. We will now discuss the two cases separately.}\\

{\bf Reheating for {HSI} ($\lambda_{H\sigma}>0$):} 
During preheating, Higgs bosons are {non-resonantly} produced during inflaton crossings because of the large value of the Higgs self-coupling \cite{Anisimov:2008qs}, as well as the fast decay of Higgses into tops and gauge bosons. {When the PQ symmetry is non-thermally restored, the} induced Higgs mass $\sqrt{\lambda_{H\sigma}}\sqrt{\langle |\sigma|^2\rangle}$ stabilises around a large value $\sqrt{\lambda_{H\sigma}} \rho_{\rm end}/a(t)\gg \omega(t)$, thus blocking Higgs production. Efficient reheating has to wait until the {spontaneously symmetry breaking (SSB)} of the PQ symmetry, i.e.\ when $\langle |\sigma|^2\rangle$ becomes $\sim v_\sigma^2$. We have simulated numerically the phase transition, finding that the energy initially stored in $\sigma$ fluctuations becomes equipartitioned into axions and $\rho$ particles. The latter can soon decay into Higgses and reheat the SM sector. The corresponding reheating temperature is $T_R\sim v_{11} \lambda_{10}^{3/8} \delta_3^{-1/8}\,10^7 \, {\rm GeV}$, where we introduce SMASH benchmark values:  $v_{11}= v_\sigma/(10^{11}$~GeV), $\lambda_{10}=10^{10}\lambda_\sigma$, $\delta_3=\delta/0.03$ \footnote{We will see that in this scenario axion DM requires $v_\sigma=f_A\sim 10^{11}$ GeV.}. The accompanying axions are relativistic and remain decoupled from such a low temperature SM thermal bath~\cite{Graf:2010tv}. They contribute to the late Universe expansion rate as extra {(relativistic)} neutrino species. We estimate $\Delta N_\nu^{\rm eff}\sim 0.96\, (\lambda_{10}/\delta_3 v_{11})^{1/6}$ above the SM value $N_\nu^{\rm eff}({\rm SM})=3.046$ \cite{Mangano:2001iu}. {Current CMB and baryon acoustic oscillation data give $N_\nu^{\rm eff} = 3.04 \pm 0.18$ at 68\% CL~\cite{Ade:2015xua}, disfavouring HSI.}\\

{\bf Reheating for {HHSI} ($\lambda_{H\sigma}<0$):} 
As in {HSI,} the direct production of Higgs excitations stops when the PQ symmetry is non-thermally restored. {However,} the Higgs component of the inflaton {continues} to oscillate around {$h\sim 0$ so that $W$ and $Z$ gauge bosons} can still be produced during crossings.  The fast decay of $W,Z$ into light fermions when $h$ moves away from zero prevents {their exponential accumulation but makes} the comoving energy in light fermions increase. 
When light particles thermalise, a population of $W,Z$ bosons is created by the thermal bath during crossings (when their mass is below the temperature) and decays when their mass grows {with $h$.} This mechanism enhances {the drain} of energy from the inflaton to the SM bath. 
Using Boltzmann equations with thermal and non-thermal sources, and accounting for the energy loss of the background {fields,} we have calculated numerically the reheating temperature, {finding $T_R\sim{\cal O}(10^{10} $GeV) for the values of $\lambda$ and $\delta$ satisfying the requirements for  inflation and stability.  

The critical temperature for the PQ phase transition is ${T_c}\simeq 2\sqrt{6\lambda_\sigma}\,v_\sigma/\sqrt{8(\lambda_\sigma+\lambda_{H\sigma})+\sum_i Y^2_{ii}+6 y^2}$ \cite{Ballesteros:2016xej}. For SMASH benchmark values $|\lambda_{H\sigma}|\gg \lambda_\sigma$, and requiring the previous stability bound on the Yukawa couplings of the new fermions, $T_c\sim 0.01\,v_\sigma< T_R$. Therefore, the PQ symmetry{, which had been non-thermally restored by preheating, is also restored thermally at the end of reheating. A few Hubble times after,} {the temperature} {drops below $T_c$ and the PQ symmetry becomes spontaneously broken, this time for good.} 
 We thus predict a thermal abundance of axions, which decouple at ${\rm min}\{T_c,T_A^{\rm dec}\}$ where $T_A^{\rm dec} \simeq 2\times 10^9\ {\rm GeV}  v_{11}^{2.246}$ \cite{Masso:2002np,Graf:2010tv,Salvio:2013iaa}. Considering $g_\ast = 427/4$ relativistic degrees of freedom at axion decoupling we get $\triangle N_\nu^{\rm eff}\simeq 0.03$, which is much smaller than in HSI and in good agreement with current data. This small value of $\triangle N_\nu^{\rm eff}$ could be probed with future CMB polarisation experiments \cite{Abazajian:2013oma,Errard:2015cxa}. As discussed in \cite{Baumann:2016wac}, a non-detection of new thermal relics with future CMB probes reaching $ \Delta N^{\rm eff}_\nu\sim 0.01$ will imply that if such relics exist they were never in thermal equilibrium with the SM.} 

{Finally, we remark that the EOS of the Universe is $w=1/3$ both in the period of inflaton oscillations in a quartic potential \cite{Shtanov:1994ce} and the non-thermally PQ restored phase because the evolution is conformal in a quartic potential. This is so both for HHSI and HSI. However, in HSI, there is a small period of matter domination before the $\rho$ particles decay to reheat the SM, whose effects on $N$ are within the uncertainties. }

\section{Dark matter}
{At the spontaneous breaking of the PQ symmetry, a network of cosmic strings is formed} {both in 
{HHSI} and HSI.} {In the first case, this happens by the standard Kibble mechanism in thermal equilibrium} {\cite{Kibble:1980mv}} {and in the second, non-thermally~\cite{Tkachev:1998dc}. }
{The evolution of the network} leads to a population of low-momentum  axions that together with those arising from the realignment mechanism \cite{Preskill:1982cy,Abbott:1982af,Dine:1982ah} constitute the dark matter in SMASH.   
Requiring that all the DM is made of axions demands
\begin{equation}
\label{classic_window_v}
3\times 10^{10}\, {\rm GeV}\lesssim v_\sigma \lesssim 1.2\times 10^{11}\, {\rm GeV},
\end{equation}
which translates into the mass window 
\begin{equation}
\label{classic_window_m}
50\, \mu{\rm eV}\lesssim m_A\lesssim 200\, \mu{\rm eV},
\end{equation}
where we have updated the results of \cite{Kawasaki:2014sqa} with the latest axion mass data~\cite{Borsanyi:2016ksw}.
The main uncertainty {arises} from the string contribution~\cite{Kawasaki:2014sqa,Fleury:2016xrz}, {which we estimate as 3-4 times larger than the misalignment one;} the uncertainty is expected to be diminished in the near future~\cite{Moore:2016itg,Fleury:2015aca}.  {The SMASH axion mass window \eq{classic_window_m} will be probed in the upcoming decade by direct detection experiments such as {MADMAX~\cite{MADMAX,TheMADMAXWorkingGroup:2016hpc}} and ORPHEUS~\cite{Rybka:2014cya}. A sizeable part of the DM in this scenario may be in the form of axion miniclusters \cite{Hogan:1988mp}, which offer interesting astrophysical signatures~\cite{Kolb:1995bu,Tinyakov:2015cgg}.}

\section{Baryogenesis}
The origin of the baryon asymmetry of the Universe  is explained in SMASH from thermal leptogenesis \cite{Fukugita:1986hr}. This requires {the} massive RH neutrinos, {$N_i$, acquiring} equilibrium abundances and then decaying when their production rates become Boltzmann suppressed. {As we have seen, in HHSI}, $T_R>T_c$ for stable models in the DM window \eqref{classic_window_m}. The RH neutrinos become massive after the PQ {SSB}, and those with masses $M_i<T_c$  retain an equilibrium abundance. The stability bound on the Yukawa couplings $Y_{ii}$ enforces $T_c > M_1$, so that at least the lightest RH neutrino stays in equilibrium. Moreover, the annihilations of the RH neutrinos tend to be suppressed with respect to their decays. This allows for vanilla leptogenesis from the decays of a single RH neutrino, which demands $M_1\gtrsim 5\times10^8$ GeV \cite{Davidson:2002qv,Buchmuller:2002rq}. However, for $v_\sigma$  as in \eqref{classic_window_v},  this is just borderline compatible with stability. Nevertheless, leptogenesis can occur with a mild resonant enhancement \cite{Pilaftsis:2003gt} for a less hierarchical RH neutrino spectrum, which relaxes the stability bound and ensures that all the RH neutrinos remain in equilibrium after the {PQ SSB.} 

\section{Future perspectives}
{
SMASH provides very clear predictions, which will be tested by the next generation of CMB, large scale structure and axion DM experiments. The model predicts a correlation between $r$, $n_s$ and a small negative value of $\alpha$, as well as tiny non-Gaussianities. It also implies the existence of a cosmic background of relativistic axions which may be detected with future CMB polarisation experiments. In SMASH, the totality of the DM in the Universe is made of cold axions with mass in the range \eq{classic_window_m}, which will be explored in the next decade. If all these features are met simultaneously, {it will be a very compelling hint in favor of SMASH}. If only one is not, the model will be ruled out. {We recall that the cosmological predictions of SMASH are reliable; as opposed to those of incomplete models such as Higgs inflation, which suffers from an early breaking of perturbative unitarity.}

SMASH provides an explanation for five of the most pressing problems in particle physics and cosmology: inflation, DM, baryogenesis, the strong CP problem  and the smallness of neutrino masses; some of which are naturalness issues. However, the model does not solve the hierarchy problem nor the cosmological constant problem. It would be interesting to explore if e.g.\ some relaxation mechanism along the lines of \cite{Abbott:1984qf,Graham:2015cka,Alberte:2016izw,Arvanitaki:2016xds} could be embedded in SMASH to solve also these problems while maintaining its minimality.
}\\

\mysections{Acknowledgments}
We thank F.\ Bezrukov, A.\ G.\ Dias, J.\ R.\ Espinosa,  D.\ Figueroa, {F.\ Finelli,} J.\ Garcia-Bellido,  {J.\ Jaeckel, F.\ Kahlh\"ofer, B.\ Kniehl, J.\ Lesgourgues,} K.\ Saikawa,  M.\ Shaposhnikov, B.\ Shuve, S.\ Sibiryakov and A.\ Westphal for discussions. The work of G.B.\ is funded by the European Union's Horizon 2020 research and innovation programme under the Marie Sk\l{}odowska-Curie grant agreement number 656794 and was partially supported by the German Science Foundation (DFG) within the Collaborative Research Center SFB 676 “Particles, Strings and the Early Universe.” G.B. thanks the DESY Theory Group and the CERN Theory Department for hospitality. 
J.R.  is  supported  by  the  Ramon  y  Cajal  Fellowship  2012-10597  and  FPA2015-
65745-P  (MINECO/FEDER). G.B. and C.T. thank the Mainz Institute for Theoretical Physics for hosting them during a workshop. C.T. thanks MIAPP for hospitality while attending a programme.



\bibliography{smashBIB}

\end{document}